\documentclass[journal]{IEEEtran}
\usepackage{amsfonts}
\usepackage{amssymb}
\usepackage{amsthm}
\usepackage{amsmath,amsfonts,amssymb}
\usepackage[dvips]{graphicx}
\usepackage{verbatim}
\usepackage{setspace}
\usepackage{bm}
\usepackage[ruled,vlined]{algorithm2e}
\usepackage{cite}
\usepackage{changepage}

\newenvironment{remark}[1][Remark]{\begin{trivlist}
\item[\hskip \labelsep {\bfseries #1}]}{\end{trivlist}}

\newcommand{\figwidth}{9}
\IEEEoverridecommandlockouts

\begin{document}

\title{Low Complexity Hybrid Precoding and Channel Estimation Based on Hierarchical Multi-Beam Search for Millimeter-Wave MIMO Systems}

\author{Zhenyu Xiao,~\IEEEmembership{Member,~IEEE,}
        Pengfei Xia,~\IEEEmembership{Senior Member,~IEEE}
and Xiang-Gen Xia,~\IEEEmembership{Fellow,~IEEE}
\thanks{This work was partially supported by the National Natural Science Foundation of China (NSFC) under grant Nos. 61571025, 61201189, 91338106, and 61231013, the Fundamental Research Funds for the Central Universities under grant Nos. YWF-14-DZXY-007, WF-14-DZXY-020 and YMF-14-DZXY-027, National Basic Research Program of China under grant No.2011CB707000, and Foundation for Innovative Research Groups of the National Natural Science Foundation of China under grant No. 61221061.}
\thanks{Z. Xiao is with the School of
Electronic and Information Engineering, Beijing Key Laboratory for Network-based Cooperative Air Traffic Management, and Beijing Laboratory for General Aviation Technology, Beihang University, Beijing 100191, P.R. China.}
\thanks{P. Xia is with the School of Electronics and Information Engineering, Tongji University, Shanghai, P.R. China.}
\thanks{X.-G. Xia is with the Department of Electrical and Computer Engineering, University of Delaware, Newark, DE 19716, USA.}

\thanks{Corresponding Author: Dr. Z. Xiao with Email: xiaozy@buaa.edu.cn.}
}

\maketitle
\begin{abstract}
In millimeter-wave (mmWave) MIMO systems, while a hybrid digital/analog precoding structure offers the potential to increase the achievable rate, it also faces the challenge of the need of a low-complexity design. In specific, the hybrid precoding may require matrix operations with a scale of antenna size, which is generally large in mmWave communication. Moreover, the channel estimation is also rather time consuming due to the large number of antennas at both Tx/Rx sides. In this paper, a low-complexity hybrid precoding and channel estimation approach is proposed. In the channel estimation phase, a hierarchical multi-beam search scheme is proposed to fast acquire $N_{\rm{S}}$ (the number of streams) multipath components (MPCs)/clusters with the highest powers. In the hybrid precoding phase, the analog and digital precodings are decoupled. The analog precoding is designed to steer along the $N_{\rm{S}}$ acquired MPCs/clusters at both Tx/Rx sides, shaping an equivalent $N_{\rm{S}}\times N_{\rm{S}}$ baseband channel, while the digital precoding performs operations in the baseband with the reduced-scale channel. Performance evaluations show that, compared with a state-of-the-art scheme, while achieving a close or even better performance when the number of radio-frequency (RF) chains or streams is small, both the computational complexity of the hybrid precoding and the time complexity of the channel estimation are greatly reduced.

%
\end{abstract}


\section{Introduction}
\IEEEPARstart{M}{illimeter-wave} (mmWave) communication is a promising technology for next-generation wireless communication owing to its abundant frequency spectrum resource, which enables a much higher capacity than the current alternatives. In fact, mmWave communication has raised increasing attention as an important candidate technology in both the next-generation wireless local area network (WLAN) \cite{Rapp_2010_60GHz_general,wang_2011_MMWCS,Park_2010_11ad,Xia_2011_60GHz_Tech,xiaozhenyu2013div} and mobile cellular communication \cite{khan_2011,alkhateeb2014mimo,choi2014coding,han2015large,roh2014millimeter,sun2014mimo,niu2015survey}. In general, mmWave communication faces the problem of high propagation loss due to the high carrier frequency. Thus, mmWave devices usually need large antenna arrays to compensate for the propagation loss. Fortunately, thanks to the short wavelength of the mmWave frequency, large antenna arrays are possible to be packed into small form factors.

Despite the possibility of using large arrays, the high power consumption of mixed signal components, as well as radio-frequency (RF) chains, makes it impractical, if not impossible, to realize a full-blown digital baseband beamforming as used in the conventional multiple-input multiple-output (MIMO) systems. In such a case, analog beamforming is considered for mmWave communication, where all the antennas share a single RF chain and generally have constant-amplitude (CA) constraint on their weights \cite{wang_2009_beam_codebook,wang_2009_beam_codebook_vtc}. As entry-wise estimation of channel status information (CSI) is time costly due to large arrays and subspace observations of the channel \cite{hur2013millimeter}, the training approach is generally adopted, including the power iteration method by exploiting the directional feature of mmWave channel \cite{Xiaozy2014BeamTrain}, and the switching beamforming which probes on a pre-defined codebook and finds the best codeword within the codebook \cite{wang_2009_beam_codebook,wang_2009_beam_codebook_vtc}. For switching beamforming, multiple-stage hierarchical search algorithms were proposed to reduce the number of measurements \cite{wang_2009_beam_codebook,wang_2009_beam_codebook_vtc,chen2011multi,Park_2010_11ad,hur2013millimeter}. These schemes first probe with low-resolution codewords, i.e. codewords with larger beam widths, and then probe with high-resolution codewords, i.e. codewords with thinner beam widths. Although these analog beamforming schemes reduce search complexity, they generally share the disadvantage of steering towards only one communication beam, i.e., these schemes are not capable of achieving multiplexing gain in addition to array gain.

In order to achieve multiplexing gain, a hybrid analog/digital precoding structure was then proposed\footnote{Beamforming in the case of multi-stream transmission is called precoding here.} \cite{alkhateeb2014mimo,choi2014coding,han2015large,roh2014millimeter,sun2014mimo,niu2015survey}, where a small number of RF chains are tied to a large antenna array. This structure enables parallel transmission, and thus provides the potential to approach the capacity bound that can be achieved via digital precoding. However, the large antenna size challenges the need of a low-complexity design of the hybrid precoding and channel estimation. In particular, the hybrid precoding may require matrix operations with a scale of antenna size, which is generally large in mmWave communication. Moreover, the channel estimation is also rather time consuming due to the large number of antennas at both Tx/Rx sides.

In \cite{alkhateeb2014channel}, an overall approach was proposed for hybrid precoding and channel estimation in mmWave MIMO systems. In the channel estimation phase, a hierarchical codebook was designed by exploiting the hybrid structure, which is different from the former ones with only analog combining \cite{wang_2009_beam_codebook,wang_2009_beam_codebook_vtc}. Based on the codebook, a hierarchical multi-beam search method was proposed to acquire $L_d$ ($L_d$ is no less than the number of streams $N_{\rm{S}}$ and is generally equal to $N_{\rm{S}}$) multipath components (MPCs)/clusters with the highest powers. With these MPCs, the channel matrix was reconstructed. In the precoding phase \cite{alkhateeb2014channel,Ayach2014}, the optimal precoding matrix is first obtained based on the estimated channel without considering the CA constraint, and then analog and digital precoding matrices are determined by minimizing the Frobenius distance between the product of the analog and digital precoding matrices and the unconstraint optimal one. By exploiting the sparse feature on the angle domain, the optimization problem is modeled to be a sparse reconstruction problem, and is solved by the orthogonal matching pursuit (OMP) approach.

Although this overall approach \cite{alkhateeb2014channel} is theoretically feasible, it may not achieve a promising performance when the number of RF chains or streams is small. Moreover, it has high computational complexity in the hybrid precoding and high time complexity in the channel estimation\footnote{The time complexity of channel estimation refers to the time slots spent in the channel estimation.}. The performance of achievable rate can be degraded by the codebook design in \cite{alkhateeb2014channel}, which depends on both the number of RF chains and $L_d$. In fact, only when both of them are large enough, good wide-beam codewords can be shaped; otherwise there may be deep sinks within the coverage of a beam, which will result in high-rate \emph{miss detection} of MPCs, and in turn poor achievable rate. Regarding the computational and time complexities, for the hybrid precoding, the singular value decomposition (SVD) and matrix multiplication of size the same as the antenna number are required, which may challenge its practical implementation, while for the channel estimation, the required number of time slots (or measurements) is proportional to $L_d^3M^2$, which is basically too high for a practical mmWave system, where $M$ is the \emph{hierarchical factor} defined in Section IV.
%

In this paper, we propose a low-complexity overall hybrid precoding and channel estimation approach. The differences between this approach and the one proposed in \cite{alkhateeb2014channel} are:

\begin{itemize}
  \item In the channel estimation phase, we propose a new hierarchical multi-beam search scheme, which uses a pre-designed analog hierarchical codebook, rather than the hybrid hierarchical codebook designed in \cite{alkhateeb2014channel}. The pre-designed analog hierarchical codebook is robust to the number of RF chains and $N_{\rm{S}}$, which guarantees a robust performance. Moreover, the proposed search scheme exploits the particular channel structure in mmWave communication and the hierarchical feature of the pre-designed codebook, which greatly reduces the required time slots.
  \item In the hybrid precoding phase, the analog and digital precodings are decoupled. The analog precoding is designed to steer along the $N_{\rm{S}}$ acquired MPCs/clusters at both Tx/Rx sides, shaping an equivalent $N_{\rm{S}}\times N_{\rm{S}}$ baseband channel, while the digital precoding operates on the $N_{\rm{S}}\times N_{\rm{S}}$ equivalent channel, which greatly lowers the operation size of the matrices.
\end{itemize}
Performance evaluations show that, compared with the approach proposed in \cite{alkhateeb2014channel}, the newly proposed approach achieves a close performance to the alternative, or even a better one when the number of RF chains or streams is small. Moreover, the computational complexity of the hybrid precoding and the time complexity of the channel estimation are greatly reduced. Specifically, antenna-size matrix operations are reduced to stream-size matrix operations for the hybrid precoding, while $L_d^3M^2$-proportional time slots are reduced to $N_{\rm{S}}M$-proportional time slots for the channel estimation.


The rest of this paper is organized as follows. In Section II,
we introduce the system and channel models, and formulate the problem.
In Sections III, we propose the hierarchical multi-beam search method for channel estimation. In Section IV, we present the hybrid precoding operation. In Section V, we show the performance evaluations. Lastly, we conclude the paper in Section VI.

\emph{Notation}: $a$, $\mathbf{a}$, $\mathbf{A}$, and $\mathcal{A}$ denote a scalar variable, a vector, a matrix, and a set, respectively. $(\cdot)^{\rm{*}}$, $(\cdot)^{\rm{T}}$ and $(\cdot)^{\rm{H}}$ denote conjugate, transpose and conjugate transpose, respectively. In addition, $[x_1,x_2,...,x_M]$ denotes a row vector with its elements being $x_i$. Some other operations used in this paper are defined as follows.
 \begin{tabbing}

~~~~\= $\mathbb{E}\{\cdot\}$~~~~~~~~\= Expectation operation.\\
 \> $|x|$  \>  Absolute value of scalar variable $x$.\\
 \> $\|\mathbf{x}\|$  \>  2-norm of vector $\mathbf{x}$.\\
  \> $\|\mathbf{x}\|_0$  \>  0-norm of vector $\mathbf{x}$.\\
  \> $\|\mathbf{X}\|_{\rm{F}}$  \>  Frobenius norm of $\mathbf{X}$.\\
  \> $[\mathbf{X}]_{i,j}$  \>  The $i$th-row and $j$th-column element of $\mathbf{X}$.\\
  \> $[\mathbf{X}]_{:,i:j}$  \>  The $i$th to $j$th columns of $\mathbf{X}$.\\
 \end{tabbing}

\section{System and Channel Models}

\subsection{System Model}

\begin{figure*}[ht]
\begin{center}
  \includegraphics[width=18 cm]{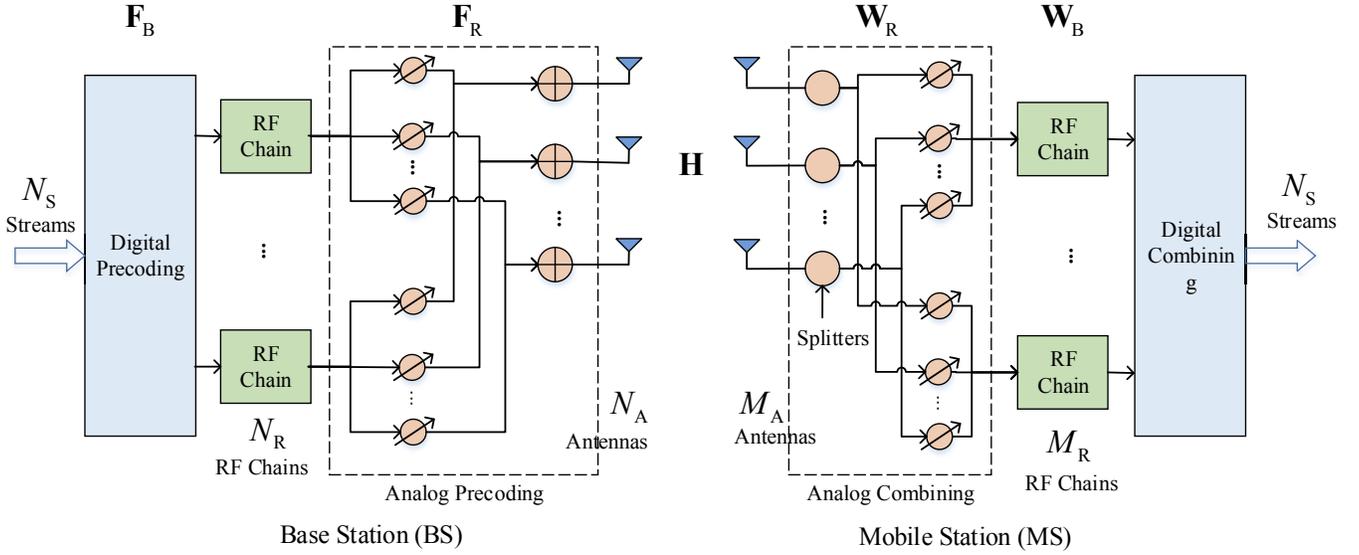}
  \caption{Illustration of the mmWave MIMO system with a hybrid analog/digital precoding and combing structure.}
  \label{fig:system}
\end{center}
\end{figure*}

Without loss of generality, we consider a downlink point-to-point multiple-stream transmission in this paper, while the signal model and the proposed scheme are also applicable for an uplink transmission. An mmWave MIMO system with a hybrid digital/analog precoding structure is shown in Fig. \ref{fig:system}, where relevant parameters are listed below.

\begin{tabbing}
 ~~~\= $N_{\rm{S}}$~~~~\=  Data streams transmitted from the base station (BS)\\
  \> \> to a mobile station (MS).\\
 \> $N_{\rm{R}}$ \>  The number of RF chains at the BS.\\
 \> $N_{\rm{A}}$ \>  The number of antennas at the BS.\\
 \> $M_{\rm{R}}$ \>  The number of RF chains at the MS.\\
 \> $M_{\rm{A}}$ \>  The number of antennas at the MS.\\
  \end{tabbing}  \vspace{-0.15 in}
Basically, we have ${N_{\rm{R}}} \le {N_{\rm{A}}}$ and $ {M_{\rm{R}}} \le {M_{\rm{A}}}$, but in practical mmWave MIMO systems, ${N_{\rm{R}}}$ and ${M_{\rm{R}}}$ are far less than ${N_{\rm{A}}}$ and ${M_{\rm{A}}}$, respectively. Moreover, it is noted that the supported number of data streams, i.e., ${N_{\rm{S}}}$, is constrained by the number of RF chains, which means ${N_{\rm{S}}} \le \min\{{N_{\rm{R}}},{M_{\rm{R}}}\}$.

 The BS performs digital precoding in the baseband and analog precoding in RF, respectively, while the MS performs analog combining in RF and digital combining in the baseband, respectively. Let ${\bf{s}}_{{N_{\rm{S}}}\times 1}$ denote the transmitted signal vector with normalized power, i.e., $\mathbb{E}({\bf{s}}{\bf{s}}^{\rm{H}})={\bf{I}}_{N_{\rm{S}}}$, where ${\bf{I}}$ is an identity matrix. Considering a narrow-band block-fading propagation channel as in \cite{Ayach2014,alkhateeb2014channel}, the received signal vector at the MS writes
\begin{equation} \label{eq_rcvSig}
{\bf{y}} = \sqrt P {\bf{W}}_{\rm{B}}^{\rm{H}}{\bf{W}}_{\rm{R}}^{\rm{H}}{\bf{H}}{{\bf{F}}_{\rm{R}}}{{\bf{F}}_{\rm{B}}}{\bf{s}} + {\bf{W}}_{\rm{B}}^{\rm{H}}{\bf{W}}_{\rm{R}}^{\rm{H}}{\bf{n}},
\end{equation}
where $P$ is the transmission power per stream, ${{\bf{F}}_{\rm{B}}}$ and ${{{\bf{F}}_{\rm{R}}}}$ are the ${{N_{\rm{R}}} \times {N_{\rm{S}}}}$ digital and ${{N_{\rm{A}}} \times {N_{\rm{R}}}}$ analog precoding matrices at the BS, respectively, ${{\bf{W}}_{\rm{B}}}$ and ${{\bf{W}}_{\rm{R}}}$ are the ${{M_{\rm{R}}} \times {N_{\rm{S}}}}$ digital and ${{M_{\rm{A}}} \times {M_{\rm{R}}}}$ analog precoding matrices at the MS, respectively, ${\bf{n}}$ is a standard white Gaussian noise vector, i.e., $\mathbb{E}({\bf{n}}{\bf{n}}^{\rm{H}})={\bf{I}}_{N_{\rm{S}}}$. In addition, we have the entry-wise CA constraint for the RF precoding matrix ${{\bf{F}}_{\rm{R}}}$ and the RF combining matrix ${{\bf{W}}_{\rm{R}}}$, respectively. In addition, we have power normalization for the precoding at the BS, i.e., $||{{\bf{F}}_{\rm{R}}}{{\bf{F}}_{\rm{B}}}||_{\rm{F}}^2 = {N_{\rm{S}}}$.

\subsection{Channel Model}
Since mmWave channels are expected to have limited scattering \cite{rapp_2011_MMW,Rapp_2012_cellular_MMW,sayeed_2007,xiao2015Iterative,alkhateeb2014channel,sun2014mimo}, MPCs are mainly generated by reflection. Different MPCs have different physical transmit steering angles and receive steering angles, i.e., physical angles of departure (AoDs) and angles of arrival (AoAs). Consequently, mmWave channels are relevant to the geometry of antenna arrays. While the algorithms and results developed in this paper can be applied to arbitrary antenna arrays, we adopt uniform linear arrays (ULAs) with a half-wavelength antenna space in this paper. Consequently, the channel can be expressed as \cite{xiao2015Iterative,he2015suboptimal,alkhateeb2014channel,Ayach2014,hur2013millimeter,nsenga_2009}
\begin{equation} \label{eq_channel}
{\bf{H}} = \sqrt {{N_{\rm{A}}}{M_{\rm{A}}}} \sum\limits_{\ell  = 1}^L {{\lambda _\ell }{\bf{g}}({M_{\rm{A}}},{\Omega _\ell }){\bf{g}}{{({N_{\rm{A}}},{\psi _\ell })}^{\rm{H}}}},
\end{equation}
where $\lambda_\ell$ is the complex coefficient of the $\ell$th path, $L$ is the number of MPCs and $L\geq N_{\rm{S}}$, ${\bf{g}}(\cdot)$ is the \emph{steering vector function}, ${\Omega _\ell }$ and ${\psi _\ell }$ are \emph{cos AoD and AoA} of the $\ell$th path, respectively. Let ${\theta _\ell }$ and ${\varphi _\ell }$ denote the \emph{physical AoD and AoA} of the $\ell$th path, respectively; then we have ${\Omega _\ell } = \cos ({\theta _\ell })$ and ${\psi _\ell } = \cos ({\varphi _\ell })$. Therefore, ${\Omega _\ell }$ and ${\psi _\ell }$ are within the range $[-1,1]$. For convenience, in the rest of this paper, ${\Omega _\ell }$ and ${\psi _\ell }$ are called AoAs and AoDs, respectively, as we design the hybrid precoding and multi-beam search schemes in the cosine angle domain. Similar to \cite{xiao2015Iterative,alkhateeb2014channel}, $\lambda_\ell$ can be modeled to be complex Gaussian distributed\footnote{Here the non-line-of-sight (NLOS) model is adopted for multi-stream transmission. For the LOS channel model in mmWave communications, the energy of the NLOS components is basically much lower than that of the LOS component \cite{rapp_2011_MMW,Rapp_2012_cellular_MMW,sayeed_2007,xiao2015Iterative,alkhateeb2014channel}; thus, one-stream transmission may be more preferable. However, we need to note that the proposed approach is also feasible in the LOS channel.}, i.e., $\lambda_\ell\thicksim \mathcal{CN}(0,1/L)$, while ${\theta _\ell }$ and ${\varphi _\ell }$ are modeled to be uniformly distributed within $[0,2\pi)$. ${\bf{g}}(\cdot)$ is a function of the number of antennas and AoD/AoA, and can be expressed as
\begin{equation}
\begin{aligned}
{\bf{g}}(N,\Omega ) =\frac{1}{{\sqrt N }}[e^ {j\pi 0\Omega},~e^{ j\pi 1\Omega },...,e^{j\pi (N - 1)\Omega}]^{\rm{T}},
\end{aligned}
\end{equation}
where $N$ is the number of antennas ($N$ is $N_{\rm{A}}$ at the BS and $M_{\rm{A}}$ at the MS), $\Omega$ is AoD or AoA. It is easy to find that ${\bf{g}}(N,\Omega )$ is a periodical function which satisfies ${\bf{g}}(N,\Omega )={\bf{g}}(N,\Omega +2)$.

%

  \begin{remark}
\textbf{1.}~Evenly sampling the cosine angle space $[-1,1]$ with an interval length $2/N$ from $-1+1/N$ leads to $N$ steering vectors ${\bf{g}}(N,-1+(2k-1)/N)$, $k=1,2,...,N$. We say each vector of them represents a \emph{basis beam} with a width $2/N$.
 \end{remark}

%

\subsection{Problem Formulation}
Our target is to maximize the total achievable rate for the downlink point-to-point transmission. Thus, the problem can be formulated as \cite[Chapters 7 and 8]{TseFundaWC}, \cite{goldsmith2003capacity}

\begin{equation} \label{eq_original_prob}
\begin{aligned}
\mathop {\max }\limits_{{{\bf{F}}_{\rm{B}}},{{\bf{F}}_{\rm{R}}},{{\bf{W}}_{\rm{B}}},{{\bf{W}}_{\rm{R}}},{\bf{Q}}} &R = {\log _2}\;\det \left( {{\bf{I}} + {\bf{K}}_{\rm{W}}^{ - 1/2}{{\bf{H}}_{\rm{E}}}{\bf{Q}}{\bf{H}}_{\rm{E}}^{\rm{H}}}{\bf{K}}_{\rm{W}}^{ - {\rm{H}}/2} \right)\\
{\rm{subject}}\;{\rm{to}}~~~~~~&||{{\bf{F}}_{\rm{R}}}{{\bf{F}}_{\rm{B}}}||_{\rm{F}}^2 = {N_{\rm{S}}},\\
&{\rm{Tr}}(\mathbf{Q})=P,
\end{aligned}
\end{equation}
where ${{\bf{H}}_{\rm{E}}} = {\bf{W}}_{\rm{B}}^{\rm{H}}{\bf{W}}_{\rm{R}}^{\rm{H}}{\bf{H}}{{\bf{F}}_{\rm{R}}}{{\bf{F}}_{\rm{B}}}$,  ${{\bf{K}}_{\rm{W}}}={\bf{W}}_{\rm{B}}^{\rm{H}}{\bf{W}}_{\rm{R}}^{\rm{H}}{{\bf{W}}_{\rm{R}}}{{\bf{W}}_{\rm{B}}}$, and $\bf{Q}$ is the power allocation matrix. To solve this problem, we must solve the following two subproblems.

\emph{Subproblem 1: Multi-Beam Search.} As we do not know the channel matrix, we need to estimate it first, which is the first subproblem. Apparently it would be rather time consuming to estimate $\bf{H}$ entry-wisely, as the antenna size is large. Fortunately, according to the structure of mmWave channel in \eqref{eq_channel}, we only need to search the AoDs and AoAs of several most significant MPCs to capture the majority of the channel energy. The search efficiency depends on the design of the hierarchical codebook and the search scheme. Different from \cite{alkhateeb2014channel}, we are going to propose an improved hierarchical multi-beam search scheme based on a pre-designed analog combining codebook.

\emph{Subproblem 2: Hybrid Precoding.} Assuming the AoAs/AoDs of the most significant MPCs have been estimated, the remaining subproblem is the hybrid precoding, i.e., to obtain ${\bf{F}}_{\rm{B}}$, ${\bf{F}}_{\rm{R}}$, ${\bf{W}}_{\rm{B}}$, ${\bf{W}}_{\rm{R}}$ and $\bf{Q}$ to maximize $R$ under the CA constraint. Note that due to the CA constraint, it is hard to find a globally optimal solution to this subproblem. On the other hand, it may be not worthy at all to find the globally optimal solution at a cost of high computational complexity, since the channel estimation is already rather approximate. Hence, we focus on designing a low-complexity hybrid precoding.

\section{Hierarchical Multi-beam Search}
In mmWave communication, channel estimation is generally realized by estimating the coefficients, AoDs and AoAs of several ($N_{\rm{S}}$ in the context of this paper) most significant MPCs, as the conventional entry-wise estimation is time consuming due to the large antenna size. Before we introduce the proposed hierarchical multi-beam search method, let us start from the introduction of the bruteforce \emph{sequential search} scheme for better understanding.


\subsection{The Sequential Search Scheme}
The sequential search scheme is straightforward, i.e., sequentially searching the whole Tx/Rx angle plane and finding the $N_{\rm{S}}$ (AoD AoA) pairs with the highest strengths. Therefore, the codebook for the sequential search consists of steering vectors with evenly sampled angles in the range of $[-1,1]$, i.e., ${\bf{g}}(N, - 1 + \frac{{2i}-1}{{KN}})$, $i=1,2,...,{{KN}}$, where the sampling interval is $\frac{2}{KN}$, and $K$ is the \emph{over-sampling factor}. The larger $K$ is, the smaller the estimation errors of AoDs and AoAs are.

Regarding the considered system in Section II, sequential search is realized by sequentially transmitting training sequences from the BS with codewords ${\bf{g}}({N_{\rm{A}}}, - 1 + \frac{{2j}-1}{{K{N_{\rm{A}}}}})$ and receiving the training sequences at the MS with codewords ${\bf{g}}({M_{\rm{A}}}, - 1 + \frac{{2i}-1}{{K{M_{\rm{A}}}}})$ for $i=1,2,...,{{K{M_{\rm{A}}}}}$ and $j=1,2,...,K{N_{\rm{A}}}$. Consequently, we can obtain the angle-domain matrix $\bf{G}$:
\begin{equation} \label{eq_ang_domain_matrix}
\begin{aligned}
&{[{\bf{G}}]_{i,j}} = {\bf{g}}{({M_{\rm{A}}}, - 1 + \frac{{2i}-1}{{K{M_{\rm{A}}}}})^{\rm{H}}}{\bf{Hg}}{({N_{\rm{A}}}, - 1 + \frac{{2j}-1}{{K{N_{\rm{A}}}}})},\\
&\quad \quad i = 1,2,...,K{M_{\rm{A}}},~j = 1,2,...,K{N_{\rm{A}}}.
\end{aligned}
\end{equation}
 Afterwards, it is straightforward to find the $N_{\rm{S}}$ most significant peaks with $\bf{G}$ on the Tx/Rx angle plane. If we denote the time period of a training sequence as a time slot (i.e., a measurement), we need $K^2{M_{\rm{A}}}{N_{\rm{A}}}$ time slots to estimate $\bf{G}$ with the sequential search approach. Considering that we have $N_{\rm{R}}$ and $M_{\rm{R}}$ RF chains at the BS and MS, respectively, in each time slot we can estimate $N_{\rm{R}}M_{\rm{R}}$ elements of $\bf{G}$, by sending different orthogonal training sequences on the $N_{\rm{R}}$ RF chains with different steering vectors at the BS, and receiving also with different steering vectors on the $M_{\rm{R}}$ chains at the MS (similar to \eqref{eq_orth_1} and \eqref{eq_orth_2}). Thus, the total time cost to estimate $\bf{G}$ is
\begin{equation}
T_{\rm{SS}}=\frac{K^2{M_{\rm{A}}}{N_{\rm{A}}}}{N_{\rm{R}}M_{\rm{R}}},
 \end{equation}
 which is proportional to ${M_{\rm{A}}}{N_{\rm{A}}}$. While this method is feasible, the time cost would be significantly high for large-array devices.

\subsection{The Hierarchical Multi-beam Search Scheme}

 \begin{figure}[t]
\begin{center}
  \includegraphics[width=\figwidth cm]{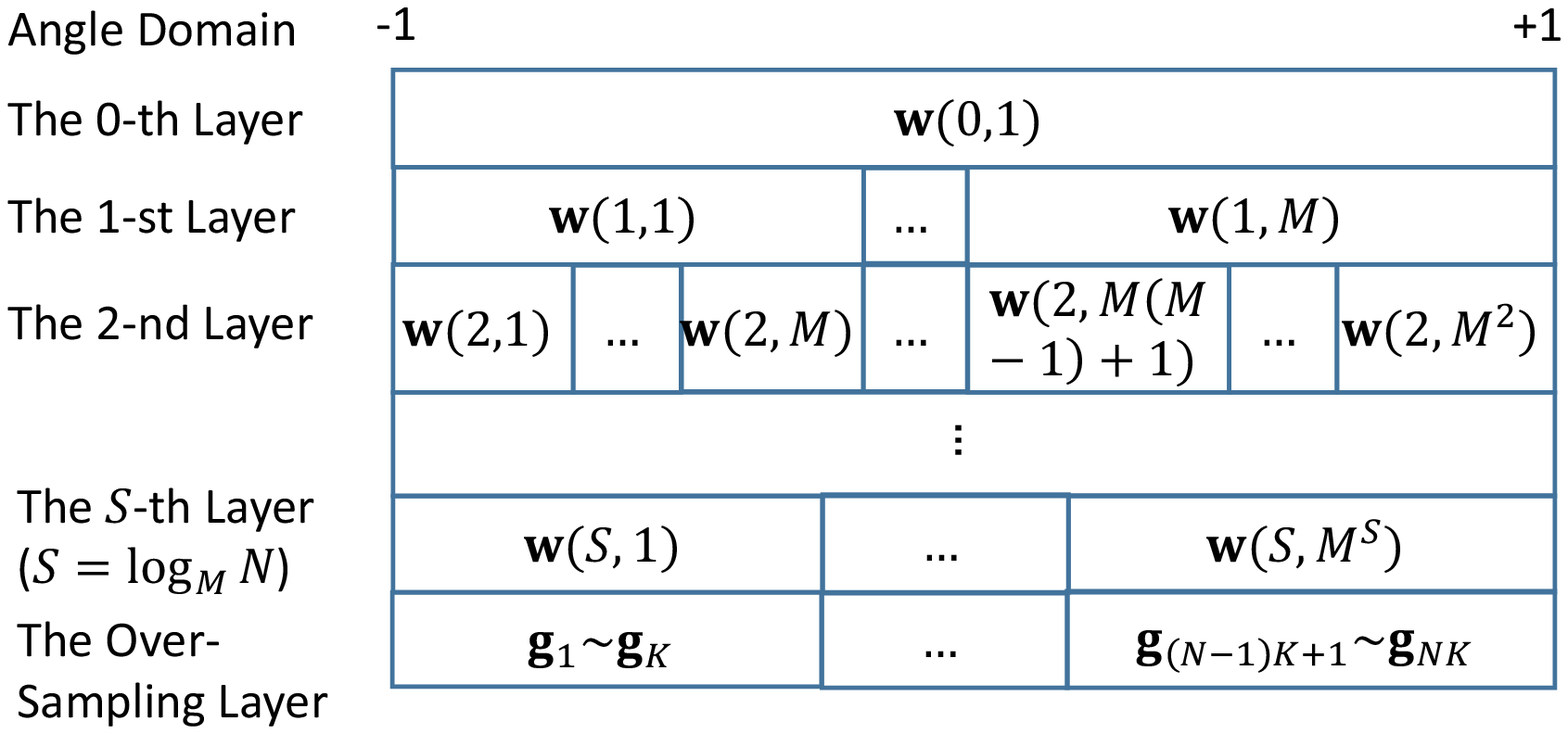}
  \caption{Structure of a hierarchical codebook.}
  \label{fig:codebook}
\end{center}
\end{figure}

To reduce the time cost for channel estimation, we propose the hierarchical multi-beam search scheme, where a corresponding hierarchical codebook needs to be designed in advance.

\subsubsection{Hierarchial Codebook Design} A typical hierarchial codebook $\mathcal{F}$ is shown in Fig. \ref{fig:codebook}, where $\mathcal{F}$ has $S+1$ \emph{layers} (not including the over-sampling layer). In the $k$th layer, there are $M^k$ codewords with the same beam width but different steering angles, $k=0,1,...,S$, where $M$ is the \emph{hierarchical factor}. The union of beam coverage of all the codewords in each layer equals $[-1,1]$. The number of antennas is assumed to satisfy $N=M^S$, where $S$ is an integer. Let ${\bf{w}}(k,n)$ denote the $n$th codeword in the $k$th layer, $n=0,1,...,M^k$. Then the beam coverage in the angle domain of ${\bf{w}}(k,n)$ is the union of those of $\{{\bf{w}}(k+1,(n-1)M+m)\}_{m=1,2,...,M}$, which is the critical feature why the codebook is \emph{hierarchical}. For convenience we say ${\bf{w}}(k,n)$ is a parent codeword of its $M$ child codewords ${\bf{w}}(k+1,(n-1)M+m)$, $m=1,2,...,M$.

There are $N$ codewords in the $S$th layer of the codebook, and they are just the evenly sampled $N$ basis beams mentioned in \textbf{Remark 1}. The angle resolution is $2/N$ in this layer. If a more accurate estimation of AoDs and AoAs is required, we need the over-sampling layer as shown in Fig. \ref{fig:codebook}, where ${{\bf{g}}_i} = {\bf{g}}(N, - 1 + \frac{{2i}-1}{KN})$, $i=1,2,...,KN$. All these $KN$ codewords within this layer are the steering vectors to sample the angle domain with an interval of $2/(KN)$. Whether or not this over-sampling layer is needed depends on practical requirements.

The design of this hierarchical codebook is challenging due to the CA constraint. In our previous work \cite{he2015suboptimal}, analog beamforming was studied, and codewords with wide beams are designed by turning off part of the antenna elements. In \cite{alkhateeb2014channel}, hybrid analog/digital beamforming was studied, and the codebook design is formulated as a sparse problem based on the hybrid structure, and solved by the OMP algorithm. Note that in \cite{he2015suboptimal} only analog combining was used for the codebook design, rather than the hybrid combining in \cite{alkhateeb2014channel}.


However, these two codebooks has separate drawbacks. For \cite{he2015suboptimal}, the number of active antennas is usually small when wide-beam codewords are configured. Because the transmit power per antenna is usually limited in mmWave communications, this would in turn limit the total transmit power in the beam search, and is undesired in general. For \cite{alkhateeb2014channel}, although multiple RF chains provides additional degrees of freedom, only when the number of RF chains is large enough, good wide-beam codewords can be generated; otherwise there will be deep sinks \cite[Fig. 5]{alkhateeb2014channel} within the beam coverage, which degrades the beam search and the achievable-rate performance.

In this paper, we choose to use the joint sub-array and deactivation (JOINT) codebook design \cite{xiao2015codebook} for hybrid beamforming. Note that in \cite{xiao2015codebook} we only proposed the codebook design, while in this paper we further consider to use it in hierarchical multi-beam search. Since the codebook with $M=2$ is widely used, we introduce the method to generate the codewords with the JOINT approach here for the case of $M=2$. However, we must note that the approach can also be used for other cases where $M$ has other values.

\emph{Codeword Generation with JOINT:}

When $k=S=\log_2(N)$, we have ${\bf{w}}(S,n)={\bf{g}}(N,-1+\frac{2n-1}{N})$, $n=1,2,...,N$.

When $k=S-\ell$, where $\ell=1,2,...,S$, we obey the following procedures to compute ${\bf{w}}(k,n)$:

\begin{itemize}
  \item {Separate ${\bf{w}}(k,1)$ into $m_S=2^{\lfloor(\ell+1)/2\rfloor}$ sub-arrays with ${{\bf{f}}_m} = {[{\bf{w}}(k,1)]_{(m - 1){n_{\rm{S}}} + 1:m{n_{\rm{S}}}}}$, where $\lfloor\cdot\rfloor$ is the flooring integer operation, $n_S=N/m_S$, $m=1,2,...,m_S$;}
  \item {Set ${\bf{f}}_m$ as \eqref{eq_codeword_setting}, where $N_{\rm{A}}=m_S/2$ if $\ell$ is odd, and $n_{\rm{A}}=M$ if $\ell$ is even;}
  \item {We have ${\bf{w}}(k,n)={\bf{w}}(k,1)\circ \sqrt N {\bf{g}}(N,\frac{2(n-1)}{N})$, where $n=2,3,...,2^k$, and $\circ$ is the entry-wise product;}
  \item {Normalize ${\bf{w}}(k,n)$.}
\end{itemize}

\begin{equation} \label{eq_codeword_setting}
{{\bf{f}}_m}=\left\{
\begin{aligned}
&e^{jm\pi}{\bf{g}}({n_{\rm{S}}},-1+\frac{2m-1}{{n_{\rm{S}}}}), ~m=1,2,...,n_{\rm{A}},\\
&{\bf{0}}_{{n_{\rm{S}}}\times 1}, ~~m=n_{\rm{A}}+1,n_{\rm{A}}+2,...,M,
\end{aligned} \right.
\end{equation}
where $n_{\rm{A}}$ is the number of active sub-arrays. See \cite{xiao2015codebook} for more details in the codebook design.

\begin{figure}[t]
\begin{center}
  \includegraphics[width=\figwidth cm]{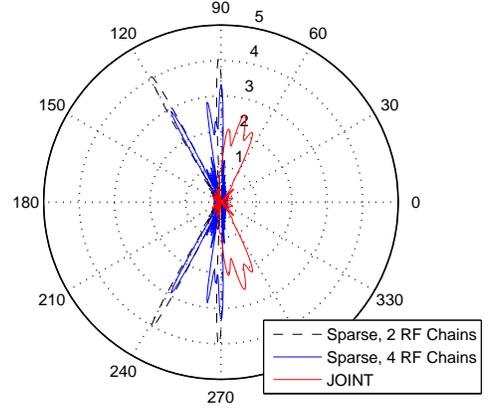}
  \caption{Comparison of beam patterns between JOINT and the approach in \cite{alkhateeb2014channel} (termed as Sparse). $N=32$, and the codewords are in the second ($k=2$) layer. $L_d=1$ for the Sparse approach.}
  \label{fig:cmp_beampattern}
\end{center}
\end{figure}

Fig. \ref{fig:cmp_beampattern} shows the comparison of the beam patterns between JOINT and the approach in \cite{alkhateeb2014channel}, where we can find that when the number of RF chains is small, there are deep sinks within the beam coverage of the wide-beam codewords, and the sink is more severe when the number of RF chains is smaller, which are in accordance with the results in \cite{alkhateeb2014channel} (Fig. 5 therein). Clearly, if the AoD or AoA of an MPC is along the sink angle, it cannot be detected with the codeword, which results in miss detection and in turn degradation of the achievable-rate performance. Besides, the beam coverage of the codewords in \cite{alkhateeb2014channel} are sensitive to both $L_d$ and the number of RF chains, i.e., different system settings need different codebooks. In contrast, there is no deep sinks within the beam coverage of the codewords in the JOINT codebook, and the JOINT approach is robust against the numbers of RF chains and data streams, because it uses analog combing.

\subsubsection{Hierarchical Multi-Beam Search} Based on the pre-designed hierarchical codebook, we next introduce the proposed beam search algorithm to find the $N_{\rm{S}}$ most significant beams on the Tx/Rx angle plane. While it is natural to search the multi-beams one by one, there are two critical issues to consider when extending a one-beam search scheme to a multi-beam search scheme.

The first issue is how to cancel the effect of the already found beams in the on-going beam search. Here we propose to exploit the particular structure of the mmWave communication channel shown in \eqref{eq_channel}, where each MPC/cluster has an AoD, AoA and path coefficient. Let

\begin{equation}
\begin{aligned}
\sqrt{P}{\bf{H}} &= \sqrt {P{N_{\rm{A}}}{M_{\rm{A}}}} \sum\limits_{\ell  = 1}^L {{\lambda _\ell }{\bf{g}}({M_{\rm{A}}},{\Omega _\ell }){\bf{g}}{{({N_{\rm{A}}},{\psi _\ell })}^{\rm{H}}}} \\
  &\triangleq \sum\limits_{\ell  = 1}^L {{\beta _\ell }{\bf{g}}({M_{\rm{A}}},{\Omega _\ell }){\bf{g}}{{({N_{\rm{A}}},{\psi _\ell })}^{\rm{H}}}}.
\end{aligned}
\end{equation}
Consequently, during the search process we can write $\bf{H}$ as
\begin{equation}
\begin{aligned}
\sqrt{P}{\bf{H}} &= \sum\limits_{\ell  \in {{\cal I}_{{\rm{fd}}}}}^{} {{\beta _\ell }{\bf{g}}({M_{\rm{A}}},{\Omega _\ell }){\bf{g}}{{({N_{\rm{A}}},{\psi _\ell })}^{\rm{H}}}}+  \\
&~~~ \sum\limits_{n \notin {{\cal I}_{{\rm{fd}}}}}^{} {{\beta _n}{\bf{g}}({M_{\rm{A}}},{\Omega _n}){\bf{g}}{{({N_{\rm{A}}},{\psi _n})}^{\rm{H}}}} \\
 &\triangleq {{\bf{H}}_{{\rm{fd}}}} + {{\bf{H}}_{{\rm{nfd}}}},
\end{aligned}
\end{equation}
where ${\cal I}_{{\rm{fd}}}$ represents the indices of the already found MPCs, while  ${{\bf{H}}_{{\rm{fd}}}}$ and ${{\bf{H}}_{{\rm{nfd}}}}$ represent the already found channel response and the not yet found channel response, respectively. During the multi-beam search, we need to perform beamforming measurements over ${{\bf{H}}_{{\rm{nfd}}}}$, which can be obtained as
\begin{equation}
\begin{aligned}
y &= {\bf{w}}_{{\rm{MS}}}^{\rm{H}}\left( {{{\bf{H}}_{{\rm{nfd}}}}{{\bf{w}}_{{\rm{BS}}}} + {\bf{n}}} \right)\\
 &= \underbrace {{\bf{w}}_{{\rm{MS}}}^{\rm{H}}\left( {\sqrt P {\bf{H}}{{\bf{w}}_{{\rm{BS}}}} + {\bf{n}}} \right)}_{{\rm{measured}}} - \underbrace { {\bf{w}}_{{\rm{MS}}}^{\rm{H}}{{\bf{H}}_{{\rm{fd}}}}{{\bf{w}}_{{\rm{BS}}}}}_{{\rm{computed}}},
\end{aligned}
\end{equation}
where the second component is just the contribution of the already found MPCs.

The second issue is how to perform layered search. According to the designed hierarchical codebook. In each layer there are $M$ candidate child codewords at both the BS and MS. A straightforward method is to exhaustively measure all the possible codeword pairs and find the best codeword pair \cite{alkhateeb2014channel}, and requires $M^2$ measurements. In the case of multi-beam search, the number of measurements becomes $M^2L_d^2$, since the number of candidate child codewords becomes $ML_d$ in each layer at both sides \cite{alkhateeb2014channel}. In our scheme we adopt a more efficient scheme, i.e., BS uses the parent codeword found in the last-layer search, while MS sequentially measures its $M$ child codewords and finds the best one; then, MS uses the found best codeword, BS sequentially measures its $M$ child codewords and finds the best one. Thus, we only need $2M$ measurements for the search in each layer.

\begin{algorithm}[htbp]\caption{Hierarchical Multi-Beam Search Algorithm.}\label{alg:coarse_search}
\textbf{1) Initialization:}

$S=\max\{\log_M {N_{\rm{A}}},\log_M {M_{\rm{A}}}\}$.\\
$i_{\rm{LY}} = 2$. $/*$The initial layer index. It can be other values depending on practical requirements.$*/$\\
${\bf{H}}_{\rm{fd}} = {\bf{0}}$, $/*$The already found channel.$*/$

\vspace{0.1 in}
\textbf{2) Iteration:}\\
\For{$\ell =1: N_{\rm{S}}$}
{
    $/*$Search for the initial Tx/Rx codewords in the $i_{\rm{LY}}$th layer.$*/$\\
    \For{$m =1: M^{i_{\rm{LY}}}$}
    {
        \For{$n =1: M^{i_{\rm{LY}}}$}
        {
            $y(m,n) = {\bf{w}}_{\rm{BS}}(i_{\rm{LY}},n)^{\rm{H}}[\sqrt{P}{\bf{H}}{\bf{w}}_{\rm{MS}}(i_{\rm{LY}},m)+{\bf{n}}]- {\bf{w}}_{\rm{BS}}(i_{\rm{LY}},n)^{\rm{H}}{\bf{H}}_{\rm{fd}}{\bf{w}}_{\rm{MS}}(i_{\rm{LY}},m)$
        }
    }
    $(m_{\rm{MS}}~m_{\rm{BS}})=\mathop{\arg \max}\limits_{(m,n)}|y(m,n)|$\\
    MS feeds back BS $m_{\rm{BS}}$.\\

    \vspace{0.1 in}
    $/*$Hierarchical refinement.$*/$\\
    \For{$s =i_{\rm{LY}}+1: S$}
    {
        \For{$n =1: M$}
        {
            $y_{\rm{MS}}(n) = {\bf{w}}_{\rm{BS}}(s-1,m_{\rm{BS}})^{\rm{H}}[\sqrt{P}{\bf{H}}{\bf{w}}_{\rm{MS}}(s,(m_{\rm{MS}}-1)M+n)+{\bf{n}}]- {\bf{w}}_{\rm{BS}}(s-1,m_{\rm{BS}})^{\rm{H}}{\bf{H}}_{\rm{fd}}{\bf{w}}_{\rm{MS}}(s,(m_{\rm{MS}}-1)M+n)$
        }
        $n_{\rm{MS}}=\mathop{\arg \max}\limits_{n}|y_{\rm{MS}}(n)|$\\
        $m_{\rm{MS}}=(m_{\rm{MS}}-1)M+n_{\rm{MS}}$\\

        \vspace{0.1 in}
        \For{$n =1: M$}
        {
            $y_{\rm{BS}}(n) = {\bf{w}}_{\rm{BS}}(s,(m_{\rm{BS}}-1)M+n)^{\rm{H}}[\sqrt{P}{\bf{H}}{\bf{w}}_{\rm{MS}}(s,m_{\rm{MS}})+{\bf{n}}]- {\bf{w}}_{\rm{BS}}(s,(m_{\rm{BS}}-1)M+n)^{\rm{H}}{\bf{H}}_{\rm{fd}}{\bf{w}}_{\rm{MS}}(s,m_{\rm{MS}})$
        }
        $n_{\rm{BS}}=\mathop{\arg \max}\limits_{n}|y_{\rm{BS}}(n)|$\\
        $m_{\rm{BS}}=(m_{\rm{BS}}-1)M+n_{\rm{BS}}$\\
        MS feeds back BS $m_{\rm{BS}}$.
    }

    \vspace{0.1 in}
    $/*$High-resolution refinement.$*/$\\
    \For{$m =(m_{\rm{MS}}-1)K+1: m_{\rm{MS}}K$}
    {
        \For{$n =(m_{\rm{BS}}-1)K+1: m_{\rm{BS}}K$}
        {
            $y(m,n) = {\bf{w}}_{\rm{BS}}(i_{\rm{LY}},n)^{\rm{H}}[\sqrt{P}{\bf{H}}{\bf{w}}_{\rm{MS}}(i_{\rm{LY}},m)+{\bf{n}}]- {\bf{w}}_{\rm{BS}}(i_{\rm{LY}},n)^{\rm{H}}{\bf{H}}_{\rm{fd}}{\bf{w}}_{\rm{MS}}(i_{\rm{LY}},m)$
        }
    }
    $(I_\ell,J_\ell)=\mathop{\arg \max}\limits_{(m,n)}|y(m,n)|$\\
    $\beta_\ell=y(I_\ell,J_\ell)$\\
    MS feeds back BS $J_\ell$.\\

    \vspace{0.1 in}
    $/*$Updating the already found channel response.$*/$\\
    ${\bf{H}}_{\rm{fd}}={\bf{H}}_{\rm{fd}}+\beta_\ell({\bf{g}}^{\rm{MS}}_{ I_\ell})({\bf{g}}^{\rm{BS}}_{ J_\ell})^{\rm{H}}$
}

\vspace{0.1 in}
\textbf{3) Result:}

    The $\ell$th ($\ell=1,2,...,N_{\rm{S}}$) index pair is $( I_\ell, J_\ell)$ within the over-sampling layer.
\end{algorithm}

The proposed hierarchical multi-beam search scheme is shown in \textbf{Algorithm 1}. There are $N_{\rm{S}}$ iterations in the search process, and a single MPC will be searched in each of them. There are three phases to search a single MPC:

 \begin{itemize}
   \item \emph{Search for the initial Tx/Rx codewords.} As in mmWave communication the transmission power is generally limited, the beamforming gain cannot be too small. Thus, the beamforming training may not start from the 0th layer, where the codeword is omni-directional and the gain is the lowest. Instead, the beamforming training may need to start from some layer, e.g., the $i_{\rm{LY}}$th layer in \textbf{Algorithm 1}, to provide sufficient start-up beamforming gain. In this process, there are $M^{i_{\rm{LY}}}$ candidate codewords at both BS and MS. Thus, an exhaustive search over all the BS/MS codeword pairs is adopted to search the best Tx/Rx codeword pair, which are treated as the parent codewords for the following search.
   \item \emph{Hierarchical refinement.} In this process, a staged search is performed to refine the beam angle step by step. The search process has $(S-i_{\rm{LY}})$ stages, and begins from the $(i_{\rm{LY}}+1)$th layer. In each stage, there are $M$ candidate codewords at the BS and MS, which are the $M$ child codewords of the found parent codeword in the last stage. At first, the BS uses the parent codeword, while the MS sequentially tests all the $M$ child codewords and find the best one. Then the MS uses the best child codeword just found, while the BS sequentially tests all the $M$ child codewords and find the best one. The found best child codewords at the BS and MS are treated as the parent codewords in the next-stage search.
   \item \emph{High-resolution refinement.} After the hierarchical refinement, the AoD and AoA of an MPC have been found with a resolution of $2/{N_{A}}$ and $2/{M_{A}}$. In practice $K$ is generally small, because a promising performance can be achieved when $K=2$, and further increasing $K$ does not significantly improve the performance, as will be shown in Section V. Hence, for each index pair $(m_{\rm{MS}},m_{\rm{BS}})$ of the codewords found in the hierarchical refinement, we just need to test all the index pairs within $[(m_{\rm{MS}}-1)K+1,m_{\rm{MS}}K]$ at the MS and $[(m_{\rm{BS}}-1)K+1,m_{\rm{BS}}K]$ at the BS, and select the one with the greatest strength as $(I_\ell,J_\ell)$. The number of required measurements is $K^2$.
 \end{itemize}

\subsection{Time Complexity Comparison}
Since there are multiple RF chains at the BS and MS, each time slot we can make multiple measurements by sending different orthogonal training sequences on different RF chains with different codewords at the BS, and receiving with different codewords at the MS. Thus, the total number of time slots for the hierarchical search method using \textbf{Algorithm 1} can be roughly computed as
\begin{equation} \label{eq_time_complexity}
\begin{aligned}
T_{\rm{HS}}&=N_{\rm{S}}\left(\Big(\log_M (N_{\rm{A}})+\log_M(M_{\rm{A}})-2i_{\rm{LY}}\Big)\left\lceil\frac{M}{N_{\rm{R}}M_{\rm{R}}}\right \rceil+\right.\\
&~~~\left.\left\lceil\frac{M^{2i_{\rm{LY}}}}{N_{\rm{R}}M_{\rm{R}}}\right \rceil+\left\lceil\frac{K^2}{N_{\rm{R}}M_{\rm{R}}}\right \rceil\right),
\end{aligned}
\end{equation}
where $\lceil\cdot \rceil$ is the ceiling integer operation, $(\log_M (N_{\rm{A}})+\log_M(M_{\rm{A}})-2i_{\rm{LY}})$ and $\lceil\frac{M}{N_{\rm{R}}M_{\rm{R}}}\rceil$ are the total number of stages and the number of measurements in each stage at the BS and MS in the hierarchical refinement, respectively, $\lceil\frac{M^{2i_{\rm{LY}}}}{N_{\rm{R}}M_{\rm{R}}}\rceil$ and $\lceil\frac{K^2}{N_{\rm{R}}M_{\rm{R}}}\rceil$ are the numbers of measurements in the initial codeword search and the high-resolution refinement, respectively. Since $\lceil\frac{M^{2i_{\rm{LY}}}}{N_{\rm{R}}M_{\rm{R}}}\rceil$ and $\lceil\frac{K^2}{N_{\rm{R}}M_{\rm{R}}} \rceil$ are irrelevant to the number of antennas, $T_{\rm{HS}}$ is roughly proportional to $N_{\rm{S}}M$.

 For a fair comparison, we assume the required number of MPCs $L_d$ is equal to $N_{\rm{S}}$. Then the required number of measurements of the scheme in \cite{alkhateeb2014channel} is\footnote{In this formula it was assumed that $N_{\rm{A}}\geq M_{\rm{A}}$.}
 \begin{equation}
 T_{\rm{SP}}=MN_{\rm{S}}^2\left\lceil \frac{MN_{\rm{S}}}{N_{\rm{R}}}\right\rceil\log_M\left(\frac{KN_{\rm{A}}}{N_{\rm{S}}}\right),
 \end{equation}
 which is roughly proportional to $N_{\rm{S}}^3M^2$.

\begin{figure}[t]
\begin{center}
  \includegraphics[width=\figwidth cm]{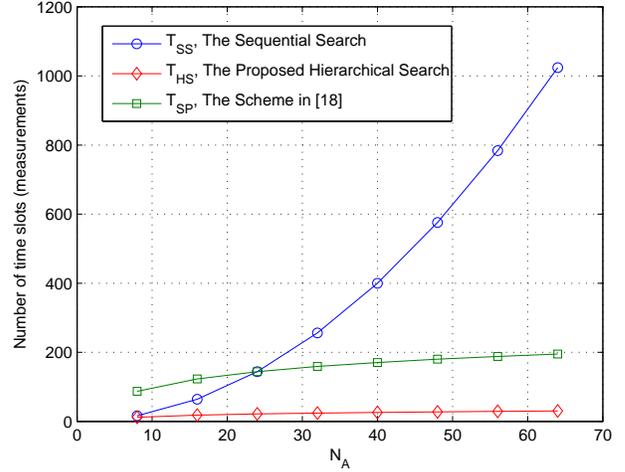}
  \caption{Comparison of time complexity between the sequential search, the proposed hierarchical multi-beam search and the scheme in \cite{alkhateeb2014channel}, where $M_{\rm{A}}=N_{\rm{A}}$, $M=2$, $K=2$, ${M_{\rm{R}}}={N_{\rm{R}}}=4$, ${N_{\rm{S}}}=3$, and $i_{\rm{LY}}=2$.}
  \label{fig:cmp_time}
\end{center}
\end{figure}

 Fig. \ref{fig:cmp_time} shows the comparison of time complexity, i.e., required time slots for beamforming training, between the sequential search, the proposed hierarchical multi-beam search and the scheme in \cite{alkhateeb2014channel}, with the parameters listed in the caption. It can be observed that the proposed hierarchical search scheme has the lowest time complexity among the three. Compared with the scheme in \cite{alkhateeb2014channel}, the proposed hierarchical search scheme achieves a further significant reduction on the required time slots.

\section{Low-Complexity Hybrid Precoding}

 With the proposed hierarchial multi-beam search approach, we can obtain the index pairs of the $N_{\rm{S}}$ strongest MPCs, i.e., $({I_1},{J_1}),({I_2},{J_2}),...,({I_{N_{\rm{S}}}},{J_{N_{\rm{S}}}})$. In this section, we perform low-complexity hybrid precoding based on the estimated channel information. In specific, we first perform analog precoding without considering the digital precoding, and then compute the digital precoding matrices in the baseband.

\subsection{Analog Precoding}
The target of analog precoding is to steer at the $N_{\rm{S}}$ most significant MPCs/clusters in the angle domain. Hence, the analog precoding and combining matrices are

\begin{equation} \label{eq_ana_matrix1}
\begin{aligned}
{{\bf{F}}_{\rm{R}}} &= \left[{\bf{g}}({N_{\rm{A}}}, - 1 + \frac{{2{J_1}}-1}{{K{N_{\rm{A}}}}}),{\bf{g}}({N_{\rm{A}}}, - 1 + \frac{{2{J_2}}-1}{{K{N_{\rm{A}}}}}),...,\right.\\
&\quad\quad\quad\quad\quad\quad\quad\quad\quad\quad\quad \left.{\bf{g}}({N_{\rm{A}}}, - 1 + \frac{{2{J_{{N_{\rm{S}}}}}}-1}{{K{N_{\rm{A}}}}}) \right],
\end{aligned}
\end{equation}
and
\begin{equation} \label{eq_ana_matrix2}
\begin{aligned}
{{\bf{W}}_{\rm{R}}} &= \left[ {\bf{g}}({M_{\rm{A}}}, - 1 + \frac{{2{I_1}}-1}{{K{M_{\rm{A}}}}}),{\bf{g}}({M_{\rm{A}}}, - 1 + \frac{{2{I_2}}-1}{{K{M_{\rm{A}}}}}),...,\right.\\
&\quad\quad\quad\quad\quad\quad\quad\quad\quad\quad\quad \left.{\bf{g}}({M_{\rm{A}}}, - 1 + \frac{{2{I_{{N_{\rm{S}}}}}}-1}{{K{M_{\rm{A}}}}}) \right],
\end{aligned}
\end{equation}
respectively.

\subsection{Digital Precoding}

While the analog precoding is to steer at the $N_{\rm{S}}$ most significant MPCs/clusters in the angle domain, the digital precoding is designed to cancel interference between different streams and perform power allocation at the BS.

Provided that ${\bf{F}}_{\rm{R}}$ and ${\bf{W}}_{\rm{R}}$ has been designed, we get an equivalent $N_{\rm{S}}\times N_{\rm{S}}$ baseband channel
\begin{equation} \label{eq_HB}
{{\bf{H}}_{\rm{B}}} = {\bf{W}}_{\rm{R}}^{\rm{H}}{\bf{H}}{{\bf{F}}_{\rm{R}}}.
\end{equation}
Thus,
\begin{equation} \label{eq_HBij}
\begin{aligned}
&{[{{\bf{H}}_{\rm{B}}}]_{i,j}} = {[{{\bf{W}}_{\rm{R}}}]_{:,i}}^{\rm{H}}{\bf{H}}{[{{\bf{F}}_{\rm{R}}}]_{:,j}}\\
 =& {\bf{g}}{({M_{\rm{A}}}, - 1 + \frac{{2{I_i}}}{{K{M_{\rm{A}}}}})^{\rm{H}}}{\bf{Hg}}({N_{\rm{A}}}, - 1 + \frac{{2{J_j}}}{{K{N_{\rm{A}}}}}).
\end{aligned}
\end{equation}
Since we have obtained $I_\ell$ and $J_\ell$, $\ell=1,2,...,L$, the estimation of ${\bf{H}}_{\rm{B}}$ can be easily realized within a single time slot as follows. Note that this can also be done in the phase of channel estimation after the multi-beam search.

 The BS transmits orthogonal training sequences ${\bf{s}}_j$ from the $j$th RF chain with codeword ${{\bf{g}}({N_{\rm{A}}}, - 1 + \frac{{2{J_j}}-1}{{K{N_{\rm{A}}}}})}$, where $j=1,2,...,{N_{\rm{S}}}$. Then the MS receives with ${N_{\rm{S}}}$ RF chains simultaneously, where ${\bf{g}}({M_{\rm{A}}}, - 1 + \frac{{2{I_i}}-1}{{K{M_{\rm{A}}}}})$ is adopted in the $i$th RF chain. Thus, at the $i$th RF chain of the MS, we observe
\begin{equation}\label{eq_orth_1}
\begin{aligned}
{{\bf{r}}_i}& = {\bf{g}}{({M_{\rm{A}}}, - 1 + \frac{{2{I_i}}-1}{{K{M_{\rm{A}}}}})^{\rm{H}}}{\bf{H}}\sum\limits_{j = 1}^{{N_{\rm{S}}}} {{\bf{g}}({N_{\rm{A}}}, - 1 + \frac{{2{J_j}}-1}{{K{N_{\rm{A}}}}})} {{\bf{s}}_j}\\
& = \sum\limits_{j = 1}^{{N_{\rm{s}}}} {{\bf{g}}{{({M_{\rm{A}}}, - 1 + \frac{{2{I_i}}-1}{{K{M_{\rm{A}}}}})}^{\rm{H}}}{\bf{Hg}}({N_{\rm{A}}}, - 1 + \frac{{2{J_j}}-1}{{K{N_{\rm{A}}}}})} {{\bf{s}}_j}.
\end{aligned}
\end{equation}

By multiplying with ${\bf{s}}_k$ at the $i$th RF chain of the MS, where $i=1,2,...,{N_{\rm{S}}}$ and $k=1,2,...,{N_{\rm{S}}}$, we get
\begin{equation}\label{eq_orth_2}
\begin{aligned}
&{\bf{s}}_k^{\rm{H}}{{\bf{r}}_i} \\
= &{\bf{s}}_k^{\rm{H}}\sum\limits_{j = 1}^{{N_{\rm{S}}}} {{\bf{g}}{{({M_{\rm{A}}}, - 1 + \frac{{2{I_i}}-1}{{K{M_{\rm{A}}}}})}^{\rm{H}}}{\bf{Hg}}({N_{\rm{A}}}, - 1 + \frac{{2{J_j}}-1}{{K{N_{\rm{A}}}}})} {{\bf{s}}_j}\\
=& {\bf{g}}{({M_{\rm{A}}}, - 1 + \frac{{2{I_i}}-1}{{K{M_{\rm{A}}}}})^{\rm{H}}}{\bf{Hg}}({N_{\rm{A}}}, - 1 + \frac{{2{J_k}}-1}{{K{N_{\rm{A}}}}})\\
= & {[{\bf{G}}]_{{I_i},{J_k}}} = {[{{\bf{H}}_{\rm{B}}}]_{i,k}},
\end{aligned}
\end{equation}
where the noise component is neglected.

After the estimation of ${\bf{H}}_{\rm{B}}$ and the analog precoding, the received signal at the MS can be rewritten as
\begin{equation}
{\bf{y}} = \sqrt P {\bf{W}}_{\rm{B}}^{\rm{H}}{{\bf{H}}_{\rm{B}}}{{\bf{F}}_{\rm{B}}}{\bf{s}} + {\bf{W}}_{\rm{B}}^{\rm{H}}{\bf{W}}_{\rm{R}}^{\rm{H}}{\bf{n}}.
\end{equation}

Let ${{\bf{R}}_{\rm{n}}} = {\bf{W}}_{\rm{R}}^{\rm{H}}{{\bf{W}}_{\rm{R}}}$, ${\bf{\tilde W}}_{\rm{B}}^{\rm{H}} = {\bf{W}}_{\rm{B}}^{\rm{H}}{\bf{R}}_{\rm{n}}^{1/2}$, and ${{{\bf{\tilde H}}}_{\rm{B}}} = {\bf{R}}_{\rm{n}}^{ - 1/2}{{\bf{H}}_{\rm{B}}}$. The received signal is equivalent to
\begin{equation}
{\bf{y}} = \sqrt P {\bf{\tilde W}}_{\rm{B}}^{\rm{H}}{{{\bf{\tilde H}}}_{\rm{B}}}{{\bf{F}}_{\rm{B}}}{\bf{s}} + {\bf{\tilde W}}_{\rm{B}}^{\rm{H}}{\bf{n}}.
\end{equation}

Regarding the digital operations, since there is no CA constraint on ${\bf{F}}_{\rm{B}}$ and ${\bf{\tilde W}}_{\rm{B}}$, they can be determined by SVD of ${{\bf{\tilde H}}_{\rm{B}}}$. Let the SVD of ${{\bf{\tilde H}}_{\rm{B}}}$ be ${{\bf{\tilde H}}_{\rm{B}}} = {{\bf{U}}_{\rm{B}}}{{\bf{D}}_{\rm{B}}}{\bf{V}}_{\rm{B}}^{\rm{H}}$, where ${{\bf{U}}_{\rm{B}}}$ and ${{\bf{V}}_{\rm{B}}}$ are left and right unitary matrices of ${{\bf{\tilde H}}_{\rm{B}}}$, and ${{\bf{D}}_{\rm{B}}}$ is a diagonal matrix with the singular values of ${{\bf{\tilde H}}_{\rm{B}}}$ listed on the diagonal in a descending order. Then we can determine the digital precoding matrices for the MS and BS immediately, i.e.,
\begin{equation}
{{\bf{W}}_{\rm{B}}} = {\bf{R}}_{\rm{n}}^{ - {\rm{H}}/2}{{\bf{U}}_{\rm{B}}},
\end{equation}
and
\begin{equation}
 {{\bf{F}}_{\rm{B}}} = {{\bf{V}}_{\rm{B}}},
  \end{equation}
respectively. Note that ${{\bf{W}}_{\rm{B}}}$ and ${{\bf{F}}_{\rm{B}}}$ require to be normalized according to the power normalization constraint on the precoding and combining matrices.

 In addition, the power allocation matrix ${\bf{Q}}$ can be generated by water-filling the total power $P$ on the $N_{\rm{S}}$ parallel channels with coefficients on the diagonal of ${{\bf{D}}_{\rm{B}}}$.

\subsection{Computational Complexity Comparison}
The proposed hybrid precoding consists of analog precoding and digital precoding. The computational complexity of the analog precoding is low, while the digital precoding requires $N_{\rm{S}}\times N_{\rm{S}}$ matrix operations, including matrix multiplication, SVD, etc. Thus, the proposed hybrid precoding scheme has an overall computational complexity $\mathcal{O}(N_{\rm{S}}^3)$ (It is known that the computational complexity of general matrix multiplication, matrix inversion, SVD are on the order of $\mathcal{O}(M^3)$, where $M$ is the matrix size \cite{Computation_Complexity}).

In contrast, the hybrid precoding approach proposed in \cite{alkhateeb2014channel} and \cite{Ayach2014} also requires matrix operations, including matrix multiplication, SVD, etc. Since it jointly designs the analog and digital precodings, the involved matrices are with size $N_{\rm{A}}\times M_{\rm{A}}$. Hence, the hybrid precoding in \cite{alkhateeb2014channel} has an overall computational complexity $\mathcal{O}(N_{\rm{A}}^3)$ (assuming $M_{\rm{A}}=N_{\rm{A}}$).

In brief, the proposed low-complexity hybrid precoding scheme reduces the computational complexity from $\mathcal{O}(N_{\rm{A}}^3)$ to $\mathcal{O}(N_{\rm{S}}^3)$, where basically $N_{\rm{S}}\ll N_{\rm{A}}$.





\section{Performance Evaluation}
In this section, we evaluate the performance of the proposed low complexity hybrid precoding (LC-HPC) based on the hierarchical multi-beam search (HIBS). The channel model introduced in Section II is adopted, where the physical angles of the MPCs are randomly generated within $[0,2\pi)$, and the average strengths of the MPCs are equal, i.e., an NLOS channel model is considered (see Footnote 3). Each performance curve is obtained by averaging $10^3$ instantaneous performances with randomly realized channel responses. In all the simulations, we set $N_{\rm{A}}=M_{\rm{A}}=32$, but we note that we have also evaluated the performances with other numbers of antennas, and similar results can be observed. In addition, the other parameters are all set typical values for mmWave communication in the simulations, e.g., the numbers of MPCs and streams (i.e., $L$ and $N_{\rm{S}}$) are basically small.

\subsection{Hierarchial Multi-Beam Search}
First, we demonstrate the performance of the HIBS scheme, for which the most critical figure of merit is the success rate to find the index pairs of multiple beams. According to \eqref{eq_channel}, there are $L$ MPCs with different AoDs and AoAs, and HIBS finds $N_{\rm{S}}$ of them. We sequentially decide whether the found $N_{\rm{S}}$ MPCs are among the original $L$ MPCs. The decision method is that if there is an $\ell$, $1\leq \ell \leq L$, satisfying $|\psi_\ell-\alpha|<1/N_{\rm{A}}$ and $|\omega_\ell-\beta|<1/M_{\rm{A}}$, where $1/N_{\rm{A}}$ and $1/M_{\rm{A}}$ are the permitted AoD and AoA errors, $\alpha$ and $\beta$ are AoD and AoA of an estimated MPC, we say this MPC is successfully searched. Only when all the $N_{\rm{S}}$ MPCs are successfully searched, the whole search process succeeds; otherwise it fails.

We note that the search of an arbitrary MPC may fail because of the noise, the effect of the previously found MPCs, and the mutual effect of MPCs, i.e., spatial fading caused by MPCs when measured with wide-beam codewords. In fact, we have simulated the performance with $L=1$ and $N_{\rm{S}}=1$, i.e., there is only one MPC. The success rate can achieve 100\% with high SNR. This is because there is no mutual effect of MPCs when $L=1$, and there is no effect of the previously found MPCs when $N_{\rm{S}}=1$.

\begin{figure}[t]
\begin{center}
  \includegraphics[width=\figwidth cm]{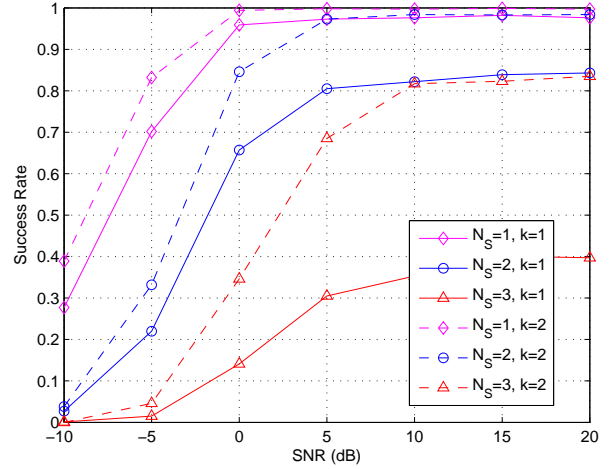}
  \caption{Success rate of the proposed hierarchical multi-beam search scheme with varying $K$, where $M=2$, $L=4$, $N_{\rm{R}}=M_{\rm{R}}=3$, $i
  _{\rm{LY}}=2$.}
  \label{fig:cmp_search1_K}
\end{center}
\end{figure}

\begin{figure}[t]
\begin{center}
  \includegraphics[width=\figwidth cm]{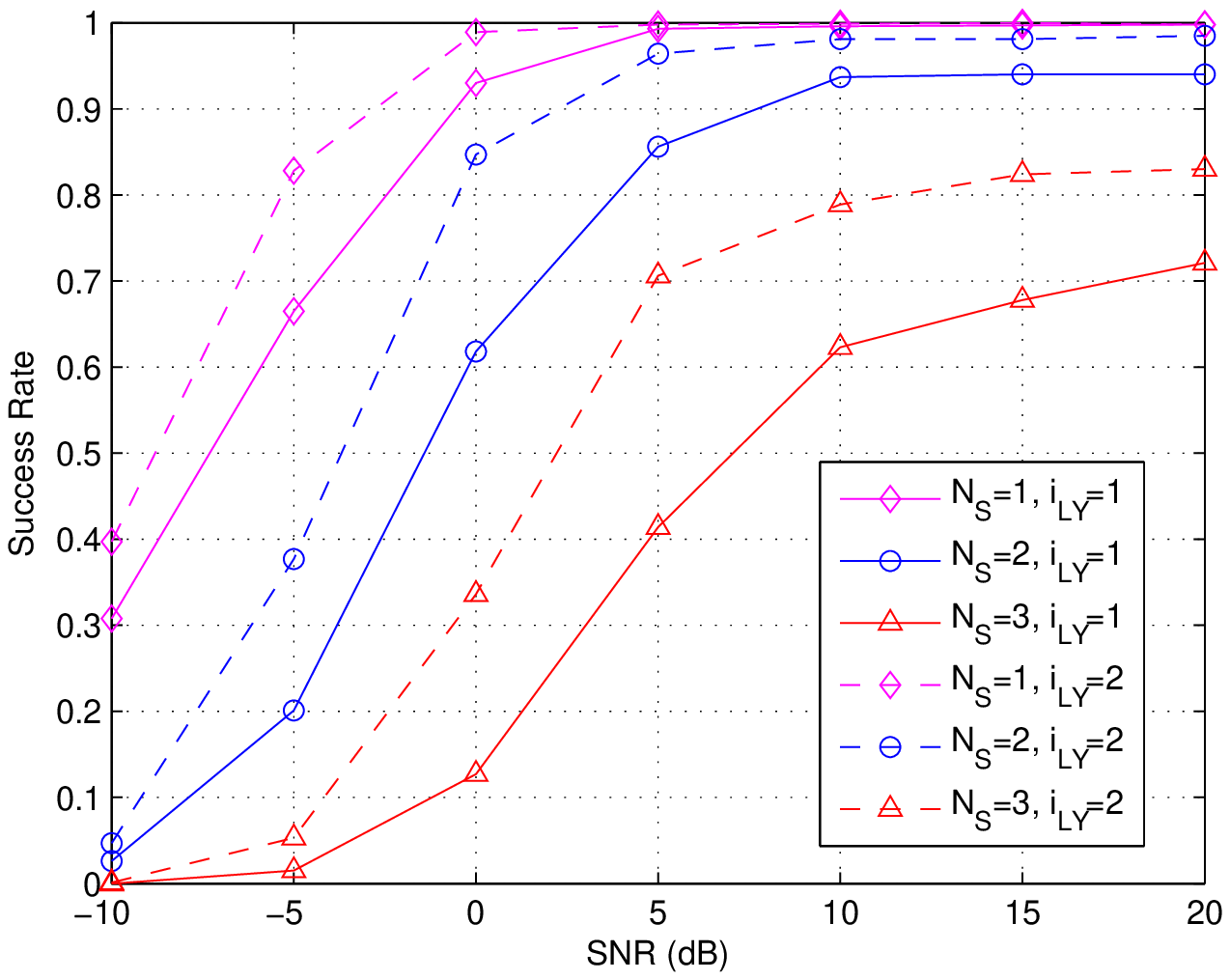}
  \caption{Success rate of the proposed hierarchical multi-beam search scheme with varying $i_{\rm{LY}}$, where $M=2$, $L=4$, $N_{\rm{R}}=M_{\rm{R}}=3$, $K=2$.}
  \label{fig:cmp_search1_iLY}
\end{center}
\end{figure}

Figs. \ref{fig:cmp_search1_K} and \ref{fig:cmp_search1_iLY} show the success rates of the proposed HIBS method with varying $K$ and $i_{\rm{LY}}$, respectively, where the SNR refers to the one after correlation operation on the training sequence. From these two figures we can find that: (i) Due to the effect of the previously found MPCs and the mutual effect of MPCs, the success rate cannot consistently increase as SNR. (ii) Basically when $N_{\rm{S}}$ is smaller, the success rate is higher. This is mainly because the contributions of the already searched MPCs cannot be completely subtracted off, and they affect the search of the next MPC. The effect becomes more significant as $N_{\rm{S}}$ increases. (iii) The success rate is higher when $K$ is larger. This is because larger $K$ means more-accurate estimation of the AoDs and AoAs, and further more-accurate contribution subtraction of the already searched MPCs. From Fig. \ref{fig:cmp_search1_K} we can find that when $N_{\rm{S}}=3$, the improvement of success rate by increasing $K$ is more significant than the cases of $N_{\rm{S}}=2$ and $N_{\rm{S}}=1$. (iv) The success rate is higher when $i_{LY}$ is bigger. Indeed, to start from a higher layer not only raises the set-up SNR, but also reduce the possible mutual effect of MPCs when measured with low-layer codewords. However, a bigger $i_{\rm{LY}}$ means a higher time complexity according to \eqref{eq_time_complexity}.

 \begin{figure*}[t]
\begin{center}
  \includegraphics[width=14 cm]{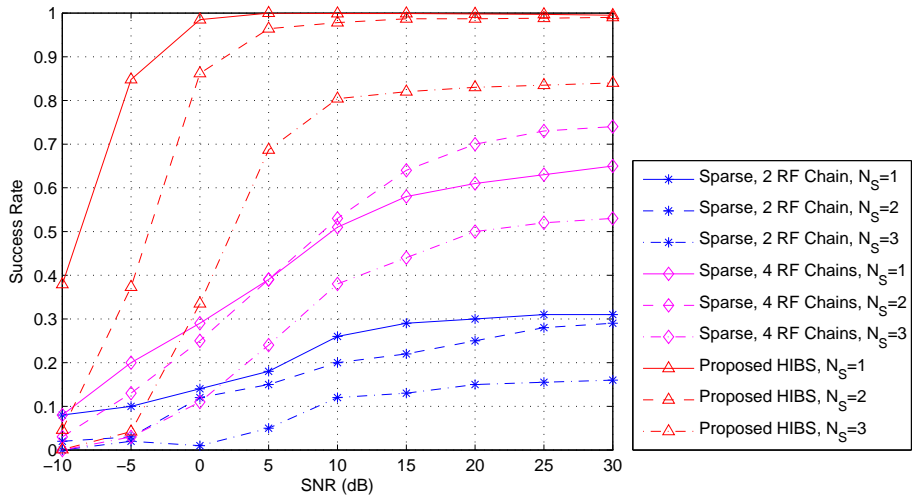}
  \caption{Comparison of success rates between the proposed hierarchical multi-beam search scheme with the search scheme in \cite{alkhateeb2014channel} (termed with ``Sparse''), where $M=2$, $L=4$, $N_{\rm{R}}=M_{\rm{R}}=3$, $K=2$, $i
  _{\rm{LY}}=2$. For the Sparse scheme, the required resolution is set as $KN_{\rm{A}}$, the same as the proposed scheme.}
  \label{fig:cmp_search1}
\end{center}
\end{figure*}

Fig. \ref{fig:cmp_search1} depicts the comparison of success rates between the proposed HIBS scheme and the search scheme in \cite{alkhateeb2014channel} (termed with ``Sparse''). From the comparison, we can find that the Sparse scheme is sensitive to the number of RF chains. Even with 4 RF chains, its performance of success rate is poorer than the proposed scheme, which requires only 1 RF chain for MPC estimation. This performance disadvantage is mainly due to the hierarchical codebook design in \cite{alkhateeb2014channel}, where the wide-beam codewords may have deep sinks within the beam coverage when the number of RF chains is not large enough, which may easily result in miss-detection of MPCs.

Moreover, from these three figures we can observe that although the desired number of MPCs is $N_{\rm{S}}$, the number of actual found MPCs with either HIBS or the Sparse scheme may be less than $N_{\rm{S}}$, especially when $N_{\rm{S}}$ is large. In such a case, the actual number of streams will be equal to the number of actual found MPCs, which is less than $N_{\rm{S}}$, resulting a degradation of the achievable rate. However, in practical mmWave communications, the number of independent MPCs/clusters is not large \cite{rapp_2011_MMW,Rapp_2012_cellular_MMW}. Thus, these two search schemes are basically suitable.

\subsection{Low-Complexity Hybrid Precoding}

Next, we evaluate the performance of achievable rate of the proposed LC-HPC scheme, and learn the effect of $K$ on the performance.

\begin{figure}[t]
\begin{center}
  \includegraphics[width=\figwidth cm]{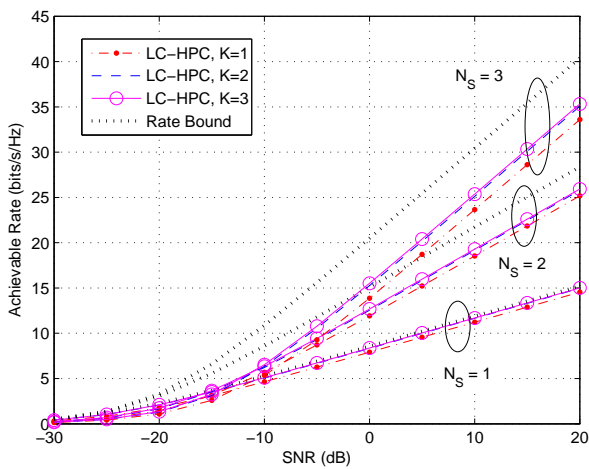}
  \caption{Achievable rate of LC-HPC with varying $K$, where the HIBS scheme is exploited to estimate MPCs. $M=2$, $L=4$, $N_{\rm{R}}=M_{\rm{R}}=3$, $i
  _{\rm{LY}}=2$.}
  \label{fig:AR_K}
\end{center}
\end{figure}

 Fig. \ref{fig:AR_K} shows the achievable rate of LC-HPC with varying $K$, where the HIBS scheme is exploited to estimate MPCs. The training sequence is assumed long enough to provide sufficiently high SNR for the MPC estimation. From this figure we find that LC-HPC achieves a promising performance. Specifically, compared with the rate bound, which is defined as the achievable rate without the CA constraint, LC-HPC has almost no loss of multiplexing gain, i.e., the slopes of the performance curves of LC-HPC are the same as those of the rate bounds. Although there is increasing SNR loss as $N_{\rm{S}}$ increases, which results from the CA constraint, it is basically acceptable in practice where $N_{\rm{S}}$ is generally small. Moreover, it is clear that the performance of LC-HPC with $K=2$ is better than that with $K=1$, but when $K\geq 2$, further increasing $K$ leads to little performance improvement. Thus, basically $K=2$ is suitable for practical usage.

\begin{figure}[t]
\begin{center}
  \includegraphics[width=\figwidth cm]{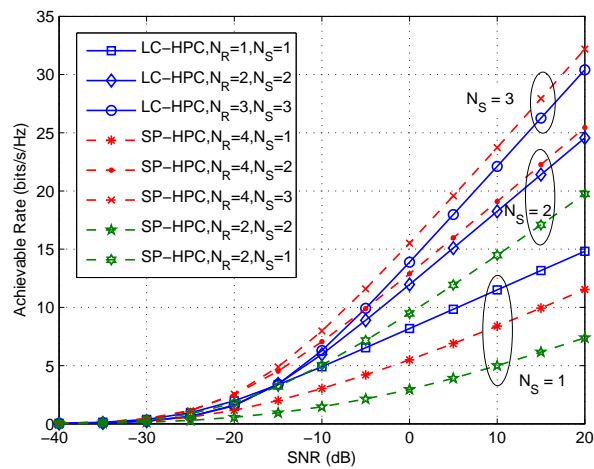}
  \caption{Comparison of achievable rates of the proposed low-complexity hybrid precoding (LC-HPC) approach with the sparse precoding approach in \cite{alkhateeb2014channel} (termed as ``SP-HPC''). $M=2$, $L=3$, $K=2$, $i
  _{\rm{LY}}=2$. $M_{\rm{R}}=N_{\rm{R}}$ for all the curves.}
  \label{fig:cmp_AR}
\end{center}
\end{figure}

 Fig. \ref{fig:cmp_AR} shows the comparison of achievable rates of the proposed LC-HPC approach with the sparse precoding approach in \cite{alkhateeb2014channel} (termed as ``SP-HPC''). The HIBS scheme is exploited for the estimation of MPCs for LC-HPC, while the search method proposed in \cite{alkhateeb2014channel} is used for the SP-HPC approach. That is to say, this figure shows the performance comparison of the overall solutions proposed in this paper and \cite{alkhateeb2014channel}. The training sequence is assumed long enough to provide sufficiently high SNR for the MPC estimations in these two approaches. From this figure we find that LC-HPC achieves a close performance to SP-HPC. The performance of SP-HPC gets improved as $N_{\rm{S}}$ and the number of RF chains increases, and is basically better than that of LC-HPC when $N_{\rm{R}}$ and $N_{\rm{S}}$ are not small. That is because SP-HPC selects the steering vectors directly from the optimization of the achievable rate, while LC-HPC just selects several estimated significant MPCs as the analog precoding matrix. However, when $N_{\rm{S}}$ or the number of RF chains is small, LC-PHC behaves even better. For instance, when $N_{\rm{S}}=1$, LC-PHC with only 1 RF chain is even better than SP-HPC with 4 RF chains. This is again due to the hierarchical codebook design in \cite{alkhateeb2014channel}, where the wide-beam codewords may have deep sinks within the beam coverage when the numbers of RF chains and $N_{\rm{S}}$ are not large enough, which may easily result in miss-detection of MPCs.

\section{Conclusions}

In this paper, a low-complexity overall hybrid precoding and channel estimation approach has been proposed. In the channel estimation phase, a new hierarchical multi-beam search scheme, which uses a pre-designed analog hierarchical codebook and the particular channel structure in mmWave communication, was proposed. While in the hybrid precoding phase, the analog precoding is designed to steer along the $N_{\rm{S}}$ acquired MPCs/clusters at both Tx/Rx sides, and the digital precoding operates on the $N_{\rm{S}}\times N_{\rm{S}}$ equivalent baseband channel.
Performance evaluations show that, compared with the approach proposed in \cite{alkhateeb2014channel}, the newly proposed approach achieves a close performance to the alternative, or even a better one when the number of RF chains or streams is small. Moreover, the computational complexity of the hybrid precoding is reduced from  $\mathcal{O}(N_{\rm{A}}^3)$ to $\mathcal{O}(N_{\rm{S}}^3)$, where basically $N_{\rm{S}}\ll N_{\rm{A}}$, while the required time slots for the multi-beam search is reduced from $N_{\rm{S}}^3M^2$-proportional to $N_{\rm{S}}M$-proportional.

\section*{Acknowledgments}
The authors would like to thank the authors of \cite{alkhateeb2014channel} to share their source code online, and particularly thank Dr. Ahmed Alkhateeb for his kind help to explain how to use the source code.




\end{document}